# Computational Intelligence based Intrusion Detection Systems for Wireless Communication and Pervasive Computing networks

Abhishek Gupta[1], Om Jee Pandey[1], Mahendra Shukla[1], Anjali Dadhich[2], Samar Mathur[3]
Dept. of ECE[1], Dept. of IT[2]
Jaipur Engineering College and Research Centre
Jaipur, India
guptaabhi3@gmail.com, anjalidadhich.it@jecrc.ac.in

Anup Ingle[4]
Dept. of E&TC[4]
Vishwakarma Institute of Information Technology[4]
Pune, India
anupingle@yahoo.com

*Abstract*— The emerging trend of ubiquitous and pervasive computing aims at embedding everyday devices such as wrist-watches, smart phones, home video systems, autofocus cameras, intelligent vehicles, musical instruments, kitchen appliances etc. with microprocessors and imparts them with wireless communication capability. This advanced computing paradigm, also known as the *Internet of Things* or cyber-physical computing, leads internet and computing to appear everywhere and anywhere using any device and location. With maximum appreciation and due regards to the evolutionary arc, depth and scope of ceaseless internet utilities, it is equally necessary to envisage the security and data confidentiality challenges posed by the free and ubiquitous availability of internet.

Wireless communication, by virtue of a plethora of networked devices, is severely prone to illegal use, unauthorized access, protocol tunneling, eavesdropping, and denial of service attacks as these devices are unknowingly exposed to illegal access from undefined locations. Amidst the rapidly expanding arena of cybercrime, banks, stock exchanges, business transactions and shopping firms such as Amazon and eBay are heavily dependent on internet. The freedom offered by wireless and 3G based internet communication and its open character has led to many incidences of abuse of technology. Unrestricted accessibility to internet has intensified the likelihood of sophisticated attacks, malicious intrusions and malware, capable of inflicting widespread damage on modern human life and economy. The classical intrusion detection systems have been found to be less equipped to handle the magnitude and complexity of wireless networks due to enormous user activities and constantly varying behavior patterns. This paper analyses the role of computational intelligence techniques to design adaptive and cognitive intrusion detection systems that can efficiently detect malicious network activities and proposes novel three-tier architecture for designing intelligent intrusion detection systems for wireless networks.

## I. Introduction

Miniaturization of electronic devices and integration of wireless communication with microprocessors has led to computers being embedded into everyday objects, breaking away from the traditional notion of desktop and laptop computers [12]. This remarkable feat of pervasive computing has been described by IBM Chairman, Mr. Lou Gerstner as the post-PC era where *a billion people interact with a million e-businesses through a trillion interconnected intelligent devices*. A vast majority of this huge number of intelligent devices comprises of popular daily use commodities such as CD players, cameras, washing machines, vehicles, smart phones etc. This transition to ubiquitous availability of internet and cyberphysical systems, though appearing quantitative, is more complex in nature, with quantitative changes of such huge orders of magnitude also implying significant qualitative changes [12].

The security of such integrated networks still remains an active area of research. The authenticity, integrity and confidentiality of information transferred over wireless network needs to be monitored and implemented effectively. Cryptography and information encryption is often considered a strong network security measurement, but they can be breached by sophisticated mathematical attacks or corrupted by bugs. As threats from internet have become progressively sophisticated, the traditional intrusion prevention mechanisms such as firewalls and other access control mechanisms have not able to render computer networks entirely safe. This has implied that these internet outages and attacks need to be detected before they could cause any damage [14]. The mechanism of detecting unauthorized access to a computer network is known as intrusion detection [10]. As the magnitude of computers has grown approximately a hundred times with introduction of pervasive and wireless networks, the traditional IDS solutions do not scale in the same way. There has been a multidimensional system development, and has rendered wireless networks prone to many old as well as new vulnerabilities [4]. The wireless security is emerging as an altogether separate domain of research because the wireless networks are inherently vulnerable to those basic security lapses from which the wired networks are easily exempted. In traditional network intrusion, the intruder specifically needs a good computer with sufficient computational power to intrude into a protected network, but with the openness of wireless networks and setting up of virtually unlimited wireless access points, the intruder can potentially access a network from anywhere without getting easily noticed or traced [3].

In this paper we have observed various security breaches that the wireless networks are prone to. To defend against such attacks and intrusions, the wireless IDS (WIDS) must work in real time and must be able to detect previously unknown attacks and extend beyond password-based user authentication methods [4]. We have analyzed the advantages offered by computational intelligence (CI) techniques that can be used to design adaptive and intelligent IDS to detect intrusions in real-time. The paper introduces novel three-tier architecture to design computationally intelligent WIDS. The rest of the paper is organized as follows: The next section highlights the IDS requirements for wireless networks, with respect to the structure of wireless networks and subtleties involved in data transfer. Section three outlines some of the most common attacks the wireless networks are prone to. The subsequent sections four and five introduce the CI paradigms and how these nature-mimicking techniques can be used to provide cognitive abilities to WIDS. Section six introduces novel three-tier architecture to represent the design of intelligent WIDS, and their communication abilities with other WIDS on the network. Finally, some conclusions are offered and future research work is stimulated.

## II. WIRELESS INTRUSION DETECTION SYSTEMS

Wireless communication need not necessarily take place only from desktop and laptop computers situated in homes and office environments. With wireless capable silicon chips getting integrated into everyday devices, internet can be accessed from any location such as a moving train, a restaurant or a park [12]. This amounts to vast and random distribution of mobile nodes that automatically self-configure without the involvement of a central authority responsible for management of infrastructure [2]. Intrusions on these systems is defined as unauthorized access to these devices via hacks and bits of malicious code, intended to either compromise with sensitive and personal user information or to deny bandwidth access to authorized users [3].

Whereas some of the network attacks and intrusions are well-funded governmental activities carried out in order to test the reliability of systems and for intelligence gathering, many other intrusions and attacks are autonomous and profit related. The attacks could be potentially destructive, effectively cutting the internet access for considerable periods, denying access to e-mails and other personal online accounts, misuse proprietary information, hacking online banking accounts, forwarding spam through e-mail servers, impeding news agencies or ceasing the ATMs to work. Over wireless networks, the magnitude of such malicious behavior increases manifold [2].

IDS designed to detect malicious activities and intrusions in wireless networks need to work in real-time to defend the user's device against intrusions, both the known as well as new kind of attacks. Instead of being deployed as separate entities, WIDS must be integrated with the clients performing data collection in cooperating with other servers and agents on the network [3]. If implemented as traditional signature-based IDS working on a set of known rules, these systems fail to respond to the environments where new intrusions evolve constantly and the network traffic profiles vary regularly [6]. WIDS require to be built around wireless agents, sensors, distributed servers and routing and management tools, as these components form the core of wireless network infrastructure, with all devices being vulnerable to intrusions. Also, with a huge number of users getting introduced to wireless devices and networks, there is a lack of exact or well-defined boundary between a legitimate activity and a malicious attack. As the wireless intruder can attack untraced and unseen from anywhere, therefore the border of defense becomes unclear and vaguely defined [13].

[13] proposed a WIDS mechanism coupled together with wireless authentication and encryption systems defined under the IEEE 802.11 protocol. This WIDS resulted in frequent false alarms and failed to adjust to the environments where new types of attacks are often unleashed and uncovered.

[7] deployed a distributed and collaborative knowledge based WIDS where each node having WIDS participates in detecting intrusion cooperatively with other nodes. The nodes were hierarchically arranged in multi-layered networks divided into various clusters with each cluster responsible for a specific task. Each WIDS at a node has a detection element and a correlation handler. Detection elements are composed of several detection components that monitor their own sub-network and generate alerts. The correlation handler transforms the alerts into a verified report of an attack. [11] proposed to extend this knowledge-based analysis to behavior-based analysis that could recognize both the legitimate user behavior as well as a behavior deviation. This behavior based WIDS was designed as a self-learning expert system that used a specific set of rules rather than explicit and pre-defined knowledge to distinguish between network activities. Due to the vast range of activities on the wireless network, the rule base soon proved to be insufficient and the WIDS led to random false alarms. Carrying these works further, we propose a computationally intelligent WIDS architecture that can adapt to the huge magnitude of wireless network activities and also discern new evolving attacks. We propose our WIDS to be built around CI techniques that provide the WIDS with cognitive as well as perception abilities to respond to network activities [7], [11], [13].

## III. WIRELESS SPECIFIC ATTACKS

The recent pervasive computing and wireless devices aimed at the home consumer market combine a network switch with a few requisite ports, an in built self-configurable access point (AP), a router and an inbuilt modem connecting to the Internet at large. The AP is connected to the switch. Primarily, attacks on wireless networks occur when an intruder eavesdrops on network traffic armed with wireless network adaptor that helps the eavesdropper to capture network traffic for analysis [13]. A part of wireless communication takes place on unlicensed public frequencies that can be used by anyone, therefore wireless network are more susceptible to attacks than wired networks. These attacks and intrusions are easy to be carried out as these attacks can be performed using just a personal computer, other computing hardware and freely available software [8]. This section outlines some common attacks observed in wireless networks. These attacks include:

*A. Access control attacks*

Access control attacks are an attempt to penetrate into a wireless network by evading weakly laid WLAN access control measures, such as 802.1X port access controls and AP filters [9]. Various types of access control attacks include.

- 802.1X RADIUS Cracking: - This attack involves recovering RADIUS keys and passwords by brute force. This attack is fairly simple to execute by installing packet capture tools on LAN paths anywhere between target AP and RADIUS server.
- Ad Hoc Associations: - Here, an attacker connects directly to an unsecured and openly available wireless station circumventing the AP security mechanisms.
- MAC Spoofing: - In this attack, the attacker reconfigures his MAC addresses to masquerade and appear as an authorized AP to legitimate users.
- Rogue Access Points: - In this attack, the attacker installs an unsecured AP inside an existing firewall; secretly creating open backdoors into otherwise protected and well-trusted networks.
- War Driving: - The attacker constantly monitors and discovers wireless LANs by listening to beacons or by continually sending probe requests, providing launch point for further sophisticated attacks [9].

*B. Confidentiality attacks*

This category of attacks imply an attempt to illegally intercept the private and confidential information sent over wireless associations, both in unencrypted as well as encrypted form using 802.11 or other encryption protocols. These attacks allow a remote attacker to recover the WEP PIN and a network's WPA/WPA2 pre-shared keys [9]. These attacks can be further classified as:

- AP Phishing: - The attacker runs a fake portal or web server on an *evil twin* AP and phishes for user logins, passwords keys and sensitive information such as bank account details and credit card numbers.
- Eavesdropping: - Hideously capturing and deciphering unprotected network traffic to gain access to potentially sensitive information.
- Evil Twin AP: - In this attack, the attacker appears masqueraded as an authorized AP and lures legitimate users into utilizing this channel by beaconing the WLAN's service set identifier (SSID). Once the users utilize this channel for data transfer, the data is captured by the attacker.
- Man in the Middle: - The most common form of confidentiality attack is running the traditional man-in-the-middle attack tools on an evil twin AP to intercept TCP sessions. MITM attacks are further used to launch session hijacking attacks where an attacker causes the user to lose his connection, and the attacker takes on the user's identity and privileges for a certain period of time. The user's system is temporarily disabled using a DoS attack or a buffer overflow exploit. Users often fail to notice the interruption as it lasts no more than a few seconds. Corporate wireless networks are set up such that the user is directed to an authentication server to attempt a connection with an AP. Once authenticated, the attacker employs the session hijacking using spoofed MAC addresses.
- WEP Key Cracking: - The attacker captures user data to construct or to recover a WEP key using active or passive methods [9].

*C. Integrity attacks*

These attacks send forged data frames over wireless networks to either mislead the recipient or to form a ground to facilitate launching another type of attack [9].

- 802.11 Data Replay: - The attacker captures 802.11 data frames and replays them after a time interval with some modifications.
- 802.11 Frame Injection: - The attacker crafts and sends forged 802.11 frames to recipients, leading them to appear on networks not designed otherwise for those users.
- 802.1X EAP Replay/RADIUS Replay: - The attacker captures 802.1X Extensible Authentication Protocols EAP/ RADIUS access messages and replays them later with mild or severe modifications [9].

*D. Authentication attacks*

Intruders make use of these attacks to steal legitimate user identities and credentials to access otherwise confidential and private networks and services [9].

- 802.1X EAP Downgrade: - In this attack, the attacker forces an 802.1X server to offer a weaker authentication measure using forged EAP response.
- 802.1X Identity Theft: - This attack involves illegally capturing the user identities from 802.1X Identity Response Packets (IRP).
- 802.1X LEAP Cracking: - This attack involves recovering user credentials from captured 802.1X Lightweight EAP (LEAP) packets using sophisticated dictionary attack tools.
- 802.1X Password Guessing: - The attacker here uses a captured identity and repeatedly attempts 802.1X authentication to guess the user's password.
- Application Login Theft: - This is the simplest authentication attack, where the attacker captures user credentials from wireless application protocols.
- Domain Login Cracking: - This attack involves recovering user credentials stored in wireless servers by cracking NetBIOS password hashes using dictionary attacks.

- Shared Key Guessing: - The attacker attempts 802.11 Shared Key Authentication using dictionary attacks or cracked WEP keys.
- VPN Login Cracking: - Intruders attack Virtual Private Networks (VPN) to recover user credentials by running brute-force attacks on VPN authentication protocols [9] [13].

*E. Availability attacks*

Availability attacks are a class of Denial of Service (DoS) attacks restricting a wireless server from providing services to authorized clients because of resource exhaustion by unauthorized clients [9].
- 802.11 Associate / Authenticate Flood: - This attack involves sending out forged authenticates or associates from random MAC addresses to fill a target AP's association table.
- 802.11 Beacon Flood: - In this attack, the attacker generates numerous counterfeit 802.11 beacons and makes it tough for stations to locate a legitimate AP.
- 802.1X EAP Length Attacks: - In this attack, the attacker sends EAP type-specific messages with illegal length fields in order to crash an AP or RADIUS server, thus denying the service.
- 802.1X EAP Failure: - In this attack, the attacker keeps track of a valid 802.1X EAP exchange, and sends a forged EAP-Failure message to that station.
- 802.1X EAP-of-Death: - Here, the attacker sends malformed 802.1X EAP Identity responses.
- 802.1X EAP Start Flood: - In this attack, the attacker floods an AP with EAP-Start messages that consume available network resources or crash the target.
- AP Theft: - This attack involves physically removing a valid AP.
- Flooding with Associations: - The data supplied by a wireless station is stored in the association table maintained by the AP. The IEEE 802.11 protocol specifies a maximum of 2007 concurrent associations to an AP. In this attack, the attacker authenticates several non-existing stations using randomly generated MAC addresses and sends a flood of spoofed associate requests so that the association table overflows and crashes.
- Forged Dissociation and Deauthentication: - In this attack, the attacker sets the source MAC address same as that of the AP. The target station, though authentic, needs to reassociate and the attacker continues to send disassociation frames for a desired period, thus leading to DoS.
- Jamming the Air Waves: - Devices such as microwave ovens and cordless phones operate on the unregulated 2.4GHz radio frequency. In this attack, an advanced attacker unleashes a large amount of noise using these devices and jams the airwaves so that the wireless LAN ceases to function.
- Queensland DoS: - In this attack, the attacker makes a channel appear busy by exploiting the communication channel, usually the CSMA/CA clear channel assessment mechanism [9].

The security mechanisms to counter intrusion attempts tend to evolve with time, but so do the methods adopted by the attackers [10]. Intrusion Detection is defined as a security technology employed to monitor individual computers or computer networks to identify abnormal activities classified as computer attacks. IDS complement conventional security mechanisms. IDS have become a basic component of every security infrastructure, and are widely implemented as a last line of defense to identify malevolent behavior. Two fundamental characteristics of IDS are *Detection Rate* (DR), which must be as high as possible, and *False Alarm Rate* (FAR), which must be low. DR is the ratio of the correctly detected attacks to total number of attacks. FAR is the ratio of number of normal or legitimate actions misclassified as attacks to the total number of legitimate actions [13]. With growing number of new attacks, IDS must be intelligent to identify previously unseen attacks. The traditional hard computing based IDS have shown high recognition rates for known attack patterns but have been of little use for unknown attacks. WIDS need to be conceptualized with an ability to learn and evolve in order to to be more accurate and efficient in face of enormous unpredictable attacks [1] [10].

## IV. COMPUTATIONAL INTELLIGENCE

A number of technological paradigms and algorithms to solve complex problems have been contrived from the studies of natural and biological systems. These algorithms are presented under the heading of Computational Intelligence. Considerable accomplishments have been actualized as a consequence of modeling biological and natural intelligence, giving rise to intelligent systems. CI makes IDS a lot easier and flexible. It has been established that conventional attack analysis and IDS design methods can be replaced with CI techniques to create more robust IDS to adapt to wireless network environments. Other than high computation speeds and adaptation to changing environments, CI offers many other significant advantages such as tolerance to imprecision, uncertainty, and approximation, low overhead, improved latency and can be made to learn preferences.

CI based systems exhibit adaptability, an ability to learn and to take care of new situations by applying reasoning, generalization, association, discovery and abstraction without relying on precise human knowledge. CI is a consortium of techniques oriented towards the analysis and design of intelligent systems. CI paradigms are basically aimed at formalizing the miraculous human ability to take rational decisions in an uncertain and imprecise environment. Whereas imprecision and uncertainty are to be avoided in traditional hard computing, these are exploited in CI to arrive at a decent solution. The most popular and frequently used CI paradigms are Artificial Neural Networks (ANN), Evolutionary Computation (EC), Artificial Immune Systems (AIS), Swarm Intelligence (SI) and Fuzzy Logic (FL) [1]. All these methods inherit inspiration from nature, as shown in Fig. 1.

| Computational Intelligence Paradigm | Derives Inspiration from | Individual constituent units | Collection of individual units forms |
|---|---|---|---|
| Artificial Neural Networks (ANN) | Biological Neural Networks | Neurons | Network |
| Evolutionary Computation (EC) | Genetic, Behavioral and Natural Evolution | Chromosomes | Population |
| Swarm Intelligence (SI) | Swarm behavior of social organisms | Ants, birds and other particles | Colonies and swarms |
| Artificial Immune Systems (AIS) | Natural Immune Systems | Immune cells and molecules | Repertories |
| Fuzzy Systems | Human Thinking Process | Rule sets | Fuzzy sets |

Fig. 1. Computational Intelligence paradigms [1]

## V. NEED OF COMPUTATIONAL INTELLIGENCE IN DESIGNING WIRELESS IDS

The fundamental reason behind using CI to design WIDS is the outright incapability of cryptography, encryption and Wireless Encryption Protocol (WEP) methods to design WIDS. Some of the WIDS implemented on wireless nodes refuse to properly connect to any WEP enabled AP/router. Even if they connect, they do not acquire an IP address unless their WEP passphrase is converted to hexadecimal format manually. As the passphrase concept is not standardized for WEP, the WEP based WIDS fail to authenticate at a number of occasions. When successful, WEP still rely on a single shared key among multiple users often leading to problems in handling compromises, and was heavily susceptible to eavesdropping. WEP is shown to be inherently flawed with various discrepancies, and has been long deprecated in favor of new protocols and standards [14].

Another reason that motivates WIDS design to be handled intelligently and in real-time is the lack of any benchmark dataset that characterizes the attack patterns observed in wireless networks. As opposed to wired networks, for which we do have the highly successful, multidimensional KDD 99 dataset that is used even today by researchers globally to test IDS, no such benchmark exists for wireless networks [13]. WIDS need to be developed as integrated components of wireless communication networks. Capable to be run continuously with minimum human supervision, WIDS need to be survivable and fault tolerant [2]. In case of a system crash, the WIDS must be able to quickly recover and resume. Rather than being tailored to a particular network, WIDS needs to constantly adapt to changes in user behavior and system over time. WIDS must recognize most of the intrusions with minimum number of false-positives. Due to huge magnitude of activities on wireless networks, WIDS need to be self-protected in case of illegal modification by an attacker [3]. CI allows for identifying user behavioral patterns and deviations from such patterns. With this strategy, we can cover a wider range of unknown attacks. Basically, intrusion detection is a security management system for computers and networks which is supposed to gather and analyze information over a network or at a device to identify possible security breaches. CI techniques provide IDS with the ability to learn the network, analyze evolving user behaviors and only report valid intrusions [1].

## VI. PROPOSED ARCHITECTURE FOR WIRELESS IDS

Intrusion detection functions include monitoring and analyzing user activities, assessing system integrity, system configurations and vulnerabilities. These tasks increase manifold when an IDS is deployed in a wireless pervasive network with a huge number of networked devices in use. We propose novel three-tier architecture to build intelligent IDS suitable for pervasive wireless networks. The proposed architecture (Fig. 2) consists of a built-in mechanism to be integrated in wireless networks and can also be deployed on individual nodes, i.e. it will be suitable to work both as Network based IDS (NIDS) as well as Host based IDS (HIDS) [8]. The first level of the architecture consists of a mechanism to locate the current Internet Protocol (IP) address of the device on which the WIDS is installed. As the IP addresses update regularly depending on the wireless network in use, the WIDS needs to keep a track of IP address that identified the device for a period of time. This is termed as location-sensing and surround-sensing, that would enable the WIDS to discern fake/forged IP addresses, in case an intruder uses them [11]. The information unit is a permanent memory in the WIDS that cognitively remembers the user behavior pattern on a particular IP address. Each WIDS is capable to store its devices' behavior pattern, and then learn neighboring or other user patterns by linking with other WIDS intelligently. Thus, an individual WIDS would not be limited to the knowledge about a particular environment but would be aware of the intense activity pattern over the internet [10].

The surround sensors in the WIDS trace the current situation of the computing device under scanner and location sensors are IP tunneling devices that trace the IP address of the source from where packets originate. A database monitors and stores all IP addresses and their patterns. The source information is important for passive safety of the wireless communication devices. The surround sensors provide information not only about the source of packets but also about their environment, i.e. a market place or a secluded place. The data retrieval unit stores the IP address and can be referred to later to interpret information such as where is attack detected, the dynamics of the source, the kind of thing is detected etc. Even though these questions look quite general and superfluous, they form a primary part of detecting attacks as this information is provided to the network behavior analysis unit. Behavior analysis and classifier units need reliable information and stable data with little preprocessing [7].

The Information Processing Unit in collaboration with SI is comprised of the network's files and is responsible for obtaining access to public records over IP addresses. Information processing forwards a detected behavior pattern to all the WIDS installed on the network using SI optimization algorithm. This saves time in training each WIDS about network behavior and thus enhances self-learning among WIDS. Data retrieval involves extracting the required information the database for further analysis in form of reports and queries [7].

The optional classifier unit and rule base would enable the WIDs to store some activities that show up occasionally; these could be subject to prolonged scrutiny before raising an alarm. The same information and the set of rules could be transmitted to other WIDs and could be dealt with in a collaborative manner using *swarm intelligence* algorithm [1]. The next block of the architecture represents CI algorithms on which the WIDS and communication chips would be based. These algorithms would facilitate cognitive behavior among the network of WIDS, with the complexity of the algorithm deciding on the number of WIDS an individual WIDS can collaborate and exchange information with. The last level of the architecture relies on the cognitive abilities provided to WIDS by CI techniques and helps to report intrusions. The clustering block clusters various sets of user behavior and activities and identifies outliers [14]. Overall, the proposed architecture integrates computational intelligence, wireless communication and intrusion detection systems on a single platform leading to faster and robust IDS for wireless communication systems.

Finally, the requirements of WIDS are developed with an integration of the three levels, with all the layers working concurrently. As a future work, the authors are looking forward to build a prototype of the model using Matlab Simulink and to define the system formally using formalization language known as Calculus of Context-aware Ambients (CCA) [5].

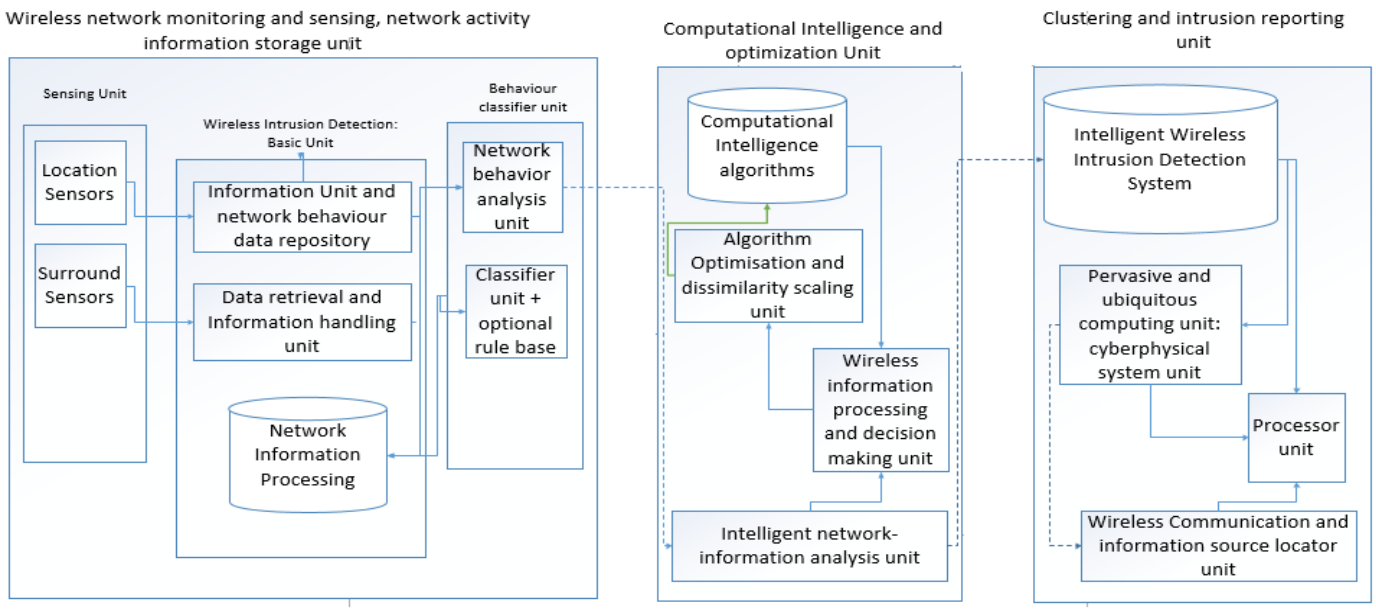

Fig. 2 Three-tier architecture with concurrently working components to design intelligent WIDS

## VII. Advantages Of The Proposed Methodology

The advantages of designing computational intelligence based wireless intrusion detection systems for pervasive and ubiquitous computing networks are enlisted below:

- The proposed WIDS is based on computational intelligence (CI) which leads to fewer errors. By virtue of the Swarm Intelligence (SI) paradigm which inherently deals with multiple particles, a number of pervasive and wireless computing systems can be connected intelligently.
- Complex intrusions and malicious activities can be detected in new environments in considerably lesser duration compared with the time it would otherwise take in the absence of intelligent detection methods.
- CI imparts a thought process to the WIDS to detect new attacks and behaviors, thus the need of a bulky rule base is avoided. SI algorithms, particularly ACO and PSO make it possible for the WIDS to take intelligent decisions on its own.
- The proposed methodology connects various wireless networks and an intrusion/malicious activity taking place on one device can be learnt by WIDS installed on all other devices on the network. This simplifies the future IDS architecture which is capable of learning, reasoning and self-analysis, i.e. capable of detecting the attacks it is not preprogrammed to perform and can discover unexplored attacks.
- The proposed method serves as an efficient complement to the previous works carried out to design IDS for various networks. Considering the large numbers of wireless and pervasive computing systems, it is a feasible and economical approach to have a policy where the IDS can learn new patterns by communication with neighboring IDS thus adding to its intelligence.

- Intrusions appear in a continuous stream in a system and each of these disturbances give rise to a chain of such activities. The proposed work gives a complete picture of the state of WIDS where each WIDS installed on internet capable devices will be able to communicate intelligently with each other.
- When WIDS networks are completely interconnected then it gets integrated into the internet architecture with less complexity. This stresses the idea that the combination of wireless communication, network security intelligence and SI is a possible phenomenon and proposed the next stage of evolution.
- The proposed method makes intelligent and faster WIDS a much simpler phenomenon.
- The approach would diminish the need of separate IDs for devices and would make IDS an inherent built-in phenomenon in pervasive and ubiquitous wireless communication systems, leading to improvement in online security measures.

## VIII. CONCLUSION AND FUTURE WORK

In this study, we have concluded that regardless of the implemented security and encryption protocols, the wireless networks continue to remain potentially insecure as the intruder can listen in without gaining physical access. Thus, the protocol designs prove to be security-naive. We have pointed out several mechanisms and tools that allow the intruder to implement attacks by exploiting the weaknesses in the protocol designs. The integration of wireless networks into existing networks also has also led to increased network vulnerabilities. The study also analyzed several practices in use to mitigate the insecurities in wireless networks and their drawbacks were studied. Major security flaws in wireless networks are also presented in the paper.

We arrive at a conclusion that amidst the evolving network environment, WIDS need to be supported with CI techniques and propose novel three-tier architecture to design intelligent WIDS. The authors are looking forward to implement the design. In the future, the authors also propose to extend the architecture to cloud computing networks. As an immediate future work, the authors look forward to implement the simulation in Matlab Simulink and formalize the system in CCA. In the future, the proposed architecture will be extended into a prototype consisting of a handful of wireless devices spread over a small area. A similar IDS architecture will be developed for cloud computing networks for data security, including middleware.

ACKNOWLEDGMENT

The authors express heartfelt gratitude to Dr. Aditya Abhyankar, Dean (Research), Vishwakarma Institute of Information Technology for motivating us to commence this project. We are thankful to Dr. Abhyankar for constantly providing us with significant insights and valuable knowledge that have been an immense source of help in carrying on the project.